\documentclass{aastex61}

\pdfoutput=1 

\usepackage{url}

\begin{document}
\title{Theory and simulation of spectral line broadening by exoplanetary atmospheric haze }
\author{Z. Felfli}
\affiliation{ITAMP, Harvard-Smithsonian Center for Astrophysics, Cambridge, MA 02138}
\affiliation{Department of Physics, Clark Atlanta University, Atlanta, GA 30314}
\author{T. Karman}
\affiliation{Joint Quantum Centre, Durham University, Durham, DH1 3LE, United Kingdom}
\author{V. Kharchenko}
\affiliation{ITAMP, Harvard-Smithsonian Center for Astrophysics, Cambridge, MA 02138}
\affiliation{Department of Physics, UConn, Storrs, CT 06269}
\author{D. Vrinceanu}
\affiliation{Department of Physics, Texas Southern University, Houston, TX 77004}
\author{J. F. Babb}
\affiliation{ITAMP, Harvard-Smithsonian Center for Astrophysics, Cambridge, MA 02138}
\author{H. R. Sadeghpour}
\affiliation{ITAMP, Harvard-Smithsonian Center for Astrophysics, Cambridge, MA 02138}

\begin{abstract}

Atmospheric haze is the leading candidate for the flattening of expolanetary spectra, as it's also an important source of opacity in the atmospheres of solar system planets, satellites, and comets. Exoplanetary transmission spectra, which carry information about how the planetary atmospheres become opaque to stellar light in transit, show broad featureless absorption in the region of wavelengths corresponding to spectral lines of sodium, potassium and water. We develop a detailed atomistic  model, describing interactions of atomic or molecular radiators with dust and atmospheric haze particulates. This model incorporates a realistic  structure of haze particulates  from small nano-size seed particles up to sub-micron irregularly shaped aggregates, accounting for both pairwise collisions between the radiator and haze perturbers,
and quasi-static mean field shift of levels in haze environments. This formalism can explain large flattening of absorption and emission spectra in haze atmospheres and shows how  the radiator - haze particle interaction affects the absorption spectral shape in the wings of spectral lines and near their centers. The theory can account for nearly all realistic structure, size and chemical composition of haze particulates and predict their influence on absorption and emission spectra in hazy environments. We illustrate the utility of the method by computing shift and broadening of the emission spectra of the sodium D line in an argon haze. The simplicity, elegance and generality of the proposed model should make it amenable to a broad community of users in astrophysics and chemistry. 

\end{abstract}


\section{Introduction}
Extrasolar planets (exoplanets) are being discovered at {a} breakneck pace, supplanting and transforming our understanding of planetary formation and habitability.  Thousands of exoplanets have been identified using a host of detection techniques, through measurements of radial velocities, transits, and lensing \citep{ExoOrg,ExoEU}.
Many of these observed planets harbor atmospheres, and will be amenable to further observations with the launch of next generation telescopes, including the James Webb Space Telescope, the Transiting Exoplanet Survey Satellite, and {with future} terrestrial telescopes, including the Giant Magellan Telescope. An Analysis of spectral compositions and parameters of exoplanetary atmospheres  is a topic of high interest, but prominent spectral features in observed transmission spectra of such exoplanets are often obscured.

Atmospheric haze is the leading candidate for the flattening of spectral transmission of expolanetary occultation, a fact that is supported by observations of solar system planets, including Earth, and cometary atmospheres \citep{zhang2017,Seager2017,Seager2010}. Radiation spectra transmitted through hazy atmospheres carry information about how these atmospheres become opaque to stellar light in transit \citep{Pont2008,Seager2017,Spake2018}. Recent laboratory experiments simulating hazy environments for super-Earths and mini-Neptunes atmospheres suggest that some of these atmospheres contain thick photochemically generated hazes \citep{Horst2018}. Because hazy environments can reflect and absorb light, their conditions ought to be explored for direct imaging of exoplanets \citep{Morley2015}.

The stellar flux occultation during a planet's transit can also give clues to the persistence of an atmosphere; the wavelength dependent measurement of the planet transit depths reveal information on atmospheric atomic and molecular composition and transparency of the atmosphere. The Na I D and K I first  
resonance lines were modeled in absorption to constrain line-of-sight atmosphere barometric parameters and cloud depths \citep{Seager2000}. Predictions made by  \citep{Seager2000} were based on the assumption that these exoplanetary atmospheres are similar to brown and cool L dwarfs with similar effective temperatures. Water is expected to be the most spectroscopically active gas, but Na I, K I, metastable He I and CO have been identified in hot Jupiter atmospheres, and in particular in HD~209458b and HD~189733b \citep{Seager2010,Sing2011,Seager2017,Spake2018,Pont2008,Pont2013}. 
Many of the detected lines are model dependent, but Na I (and K I) resonant doublets are not, as there are no other absorbers at such wavelengths. 

In methane dwarfs \citep{Burrows2000,BurrowsRMP2001}, the K I doublet $4 ^2S-4 ^2P$ absorption line at 769 nm is  broad and is responsible for large continuum depression, red shifted in optical spectra. The continuum broadening is induced due to collisional interaction (pressure broadening) between  the radiator and background gas atoms and molecules. This lower than expected contrast of the strength of the alkali metal lines \citep{Fortney2003} has been puzzling: subsolar elemental abundances, stellar radiation ionization, and atmospheric haze particulates have been invoked as sources for the diminished strength of the lines. 

The haze hypothesis has gained currency \citep{Burrows2014,Seager2017}. Spectra of radiation observed  from different planetary and exoplanetary atmospheres, comets, and natural satellites  reveal unusual lack of sharp spectral
features,  attributed  to the presence of atmospheric dust, ice and haze particles \citep{Wong2003,Greenberg1999, West1991,Ortiz1996, Rannou2002,West1991,Tomasko1986}. Perhaps the most compelling evidence for the direct role played by haze layers in flattening of the exoplanetary lines is the transit spectra in near-UV and mid-IR by \citet{Pont2013,Pont2008} of HD189733b. Lack of features in the range from 550 to 1050 nm in the spectra of HD~189733b is suggestive of absorption by condensates high in the atmosphere \citep{Pont2008}; a strong indication of presence of sub-micron haze clusters. In general, hazes are  clouds of {small (sub-micron) particulates which can produce featureless continuum opacity to light. The red hue of Jupiter is likely produced by unidentified trace species (or haze) \citep{Burrows2014}. The transit spectra for the mini-Neptune GJ~1214b have been shown by \citep{Kreidberg2014} to be 5-10 times flatter than a water-rich, H$_2$-dominated atmosphere with, heralding the presence of a thick haze layer. {\it In situ} measurements of {the effective density of aerosol materials} in analogue Titan atmospheres confirm densities of haze-particle materials  around 0.4-1.13 g/cm$^{3}$, which for a methane dominated hazy environment, translates to the atom/molecule number  densities of a few times 10$^{22}$ cm$^{-3}$ \citep{Horst2013}.

We present a theoretical framework for a quantum mechanical description of the shift and broadening of atomic and molecular spectral lines by dust and haze particles.
Our model can be applied to shift and broadening analysis of different lines of atomic and molecular spectra,  when haze particulates may have liquid or solid structures or consist of high porosity materials.  Spectral  calculations  are obtained through the introduction of the pseudo-potential  for the interaction between radiator electrons  and  atoms and molecules of haze particles \citep{Szasz1985}. The pseudo-potential method  allows us to carry  out calculations for  different  chemical compositions of haze \citep{Jacquet2011,Boatz1994}. The binary collision between the radiator and haze particulate, affecting the radiator line 
center, is also considered \citep{Allard1982}.

\section{Methodology}

\subsection{Modeling of Emission  and Absorption Spectra in Haze Environments}
 The physical nature, composition and structure of these particles 
is roughly  known  for the Earth, some  solar system planets, satellites, and comets  \citep{Greenberg1999,West1991,Seignovert2017}, but  
for exoplanetary  atmospheres {the nature of the particles is not known} \citep{Tinetti2007}.
Analysis and interpretation of  atomic and molecular spectra   observed from  dusty environments is  a formidable  task because  of complicated  quantum mechanical  interaction  between  radiating atoms/molecules   and haze particles.

Another fundamental obstacle in modeling optical properties of mesoscopic dust, ice and haze particles is the stochastic  nature of the haze particle distribution \citep{Greenberg1999,Sciamma2017}. The simplest approaches 
to modeling distributions employ empirically  assigned refractive indices in the classical Mie model for scattering and absorption of radiation by spherical haze particles \citep{Lenoble2013,Seignovert2017}.  This model fails to describe  realistic  changes in atomic and molecular emission/absorption spectra because optical  properties of nano-size  particles cannot be described by {macroscopic parameters such as the} refractive index.

\subsection{Quantum Mechanical Model of Interaction Between Radiator and Haze Particles }
The electronic Hamiltonian ${H}_{el}$ for the radiator-haze {system} can be written as:
\begin{equation}
{H}_{el} ={H}_r + {H}_{hp} + V_{r-hp},
\end{equation}
where  ${H}_r$ is an electronic Hamiltonian of radiating atom or molecule; ${H}_{hp}$ the haze {particle} Hamiltonian, and $V_{r-hp}$  the potential of interaction between the haze and radiator electrons. The haze {particle} Hamiltonian $\hat{H}_{hp}$  includes  kinetic energies and interaction potential {energies} of all atomic and molecular constituents.

The term $V_{r-hp} $ {is given by}  the sum of interaction potential {energies} $V_i$ between radiator electrons and an atomic or molecular constituent in the haze at location $\bf{R}_{i} $ : $V_{r-hp} = \sum_{i=1}^{N} V_{i}(\bf{r}_{e},\bf{R}_{r},\bf{R}_{i}) $,  where $N$ is the total number of haze atoms and/{or} molecules, $\bf{r}_{e}$ is the electron coordinate from the radiator center $\bf{R}_{r}$. Figure \ref{diagram} illustrates the basic configuration of the haze-radiator system.

\begin{figure}[ht]
\centerline{\includegraphics[width=4.0in]{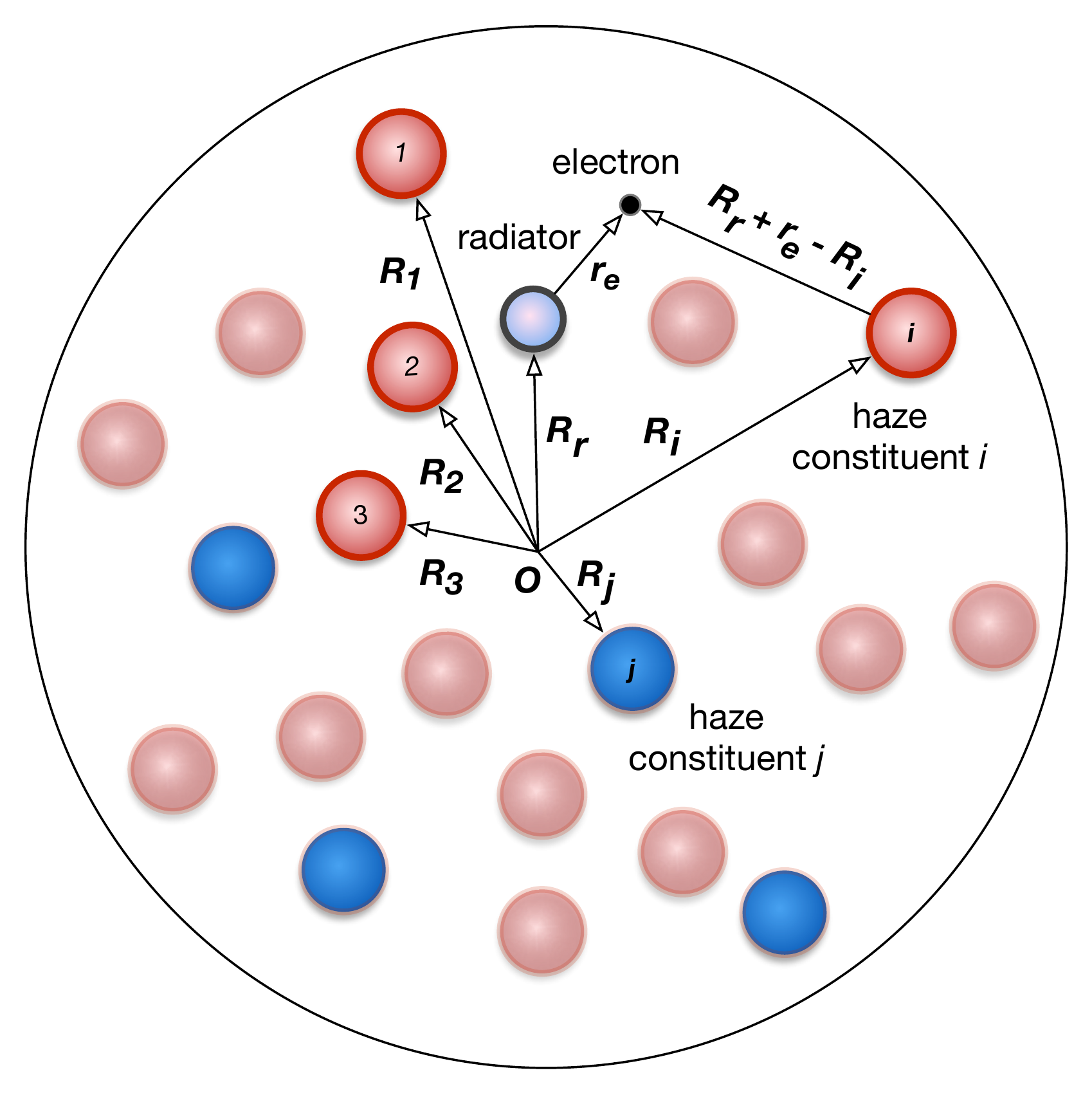}}
\caption{The active radiator electron, $\bf{r}_{e}$, interacts with the radiator core, $\bf{R}_{r}$, and with all other atoms within the haze at positions $\bf{R}_{i}$. Haze constituents of different chemical compositions are denoted with red and blue spheres. }
\label{diagram} 
\end{figure}

There are only dozens of atoms or molecules in nano-size seed particles and millions in sub-micron haze aggregates. The set of location vectors $ \{\bf{R}_{i} \} $  determines the geometrical structure of haze, its physical properties and characteristics of  interaction with radiator. For mesoscopic size ``haze-radiator'' system, the set $ \{\bf{R}_{i} \} $ has typically an irregular structure and resembles an amorphous or liquid cluster.
Nevertheless, seed particles could be represented by nano-crystals or amorphous Si or C materials covered by  polycrystalline structures of  H$_2$O and CO$_2$   ices  or  CH$_4 $ droplets  expected as  outer shells in highly porous  sub-micron haze aggregates.

\subsubsection{Mean-field spectral line shift  in haze environment}
The atmospheric distribution of haze particles is non-uniform as is the distribution of atoms and molecules inside haze droplets. Our modeling of spectral line shift and broadening is focused on the quantum mechanical description of the interactions between the haze particles and the atomic or molecular radiators. {The many-body interaction between the emitter and haze constituents is considered in the mean-field approximation.
 
The Fermi pseudopotential method \citep{Fermi1934} was first developed to describe species-dependent mean field pressure shift and broadening of Rydberg lines in gaseous environments and more recently in the context of Rydberg molecule formation in ultracold quantum gases \citep{Greene2000}.

The Fermi pseudopotential is defined is proportional to the scattering length $L_i$ of the radiator {electron} from the $i$-th haze perturber \citep{Fermi1934}, see Fig.~\ref{diagram},
\begin{equation}
V_{i }({\bf r}_{e},{\bf R}_{r},{\bf R}_{i})  = \frac{2 \pi\hbar^2}{m_e}  L_{i} \delta( {\bf R}_{r}+{\bf r}_{e} - {\bf R}_{i})
\end{equation}
where $\bf{r}_e $ is the radiator electron coordinate, $L_{i}$ is the scattering length of the electron of mass $m_e$ colliding with the haze atom/molecule at ${\bf R}_{i}$, $\hbar$ is the reduced Planck constant, and $\delta()$ is the Dirac delta-function. The chemical composition of haze particles is represented by a set of individual scattering lengths $\{L_i\}$. The scattering length is obtained from the zero-energy limit of the scattering phase shift.  Presence of  perturbers in the vicinity of radiator modifies  the radiator wave function and energy spectra that finally leads to changes in energy and rates of emitted/absorbed photons.

{Models of spectral shift and broadening \citep{Allard1982,Szudy1996} describe  the influence of collisional or {quasi-static} interaction potentials on emission and absorption spectra.  In the quasi-static approach, the line shift {$\Delta\omega = \omega - \omega_0$} induced in binary interaction is equal to the difference of radiator electron energies as a function of the perturber-radiator distance, i.e. $\hbar\Delta\omega = \Delta V(R)$, where  $\Delta V(R)$ is the difference of the ground and excited Born-Oppenheimer potential energies, 
e.g. Sec.~\ref{fig:PEC}, induced by the perturber. The inverted equation $R_{C}= R(\Delta \omega) $ yields the location of Condon points, resulting in specific frequency shifts $\Delta \omega$}. For the simplest case of cold and dilute gas of uniformly distributed perturbers, the intensity of the spectral line is proportional to the probability to find a perturber at distances between $R$ and $R+dR$, i. e. $I(\Delta\omega) = 4\pi\rho R^2(\Delta\omega) |\frac{dR(\Delta\omega)}{d(\Delta\omega)}|$, with $\rho$ the perturber number density \citep{Allard1982}.  Haze particles and aggregates may be considered as an environment of slow perturbers with a large and essentially  non-uniform  perturber number density $\rho(R)$.

The  mean-field model in this work, is also applicable to {calculations} of collisional shift and broadening inside haze aggregate particles with high levels of porosity ($\sim $90\%). Collisional and mean field mechanisms   dominate in different regions of  emission and absorption spectra. Interaction between  haze particles and radiator  can provide  significant shifts of  atomic and  molecular  lines, in extreme cases  up to $10^4$ cm$^{-1}$ \citep{Jacquet2011} depending  on the   radiator position. 

Analytical formulas for the shifted electron energies $\epsilon_{i,f}({\bf R}_{r})$  of a radiator in the ground (initial) and excited (final) states obtained from the leading term of the perturbation theory:
\begin{equation}
  \epsilon_{i,f}({\bf R}_{r})=  \epsilon^{0}_{i,f} +  \langle\Psi_{i,f}({\bf r}_e)|V_{i,f}|\Psi_{i,f}({\bf r}_e) \rangle = \epsilon^{0}_{i,f} + \frac{2\pi\hbar^2}{m_e}  \sum_{i=1}^{N} {L_{i}} |\Psi_{i,f}({\bf r}_e = {\bf R}_{i}-{\bf R}_{r})|^2
  \label{eq:PS}
  \end{equation}
where $\Psi_{i}$ and $\Psi_{f}$ are wave functions of the ground and excited states of  free (unperturbed) radiator atoms  and $\epsilon_{i,f}^0$ are related electronic energies. For systems with high densities of perturbers, such as haze ice, dust or liquid droplets, the perturbation can be strong and radiator electronic wave functions need be renormalized according to the mean field induced by the perturbers  \citep{Demkov1988}. For haze particles and {aggregates} with a large number of equivalent atoms or molecules $N$,  the summation in Eq.~\ref{eq:PS} can be replaced with integration over the volume of the haze aggregate with  a specific volume density { $\rho (\bf{R})$. For a collection of $N$ identical perturbers, the Born-Oppenheimer electronic energies of the radiator $\epsilon_{i,f}$   for initial $i$- and final $f$-states are calculated as:
\begin{equation}
\epsilon_{i,f}({\bf R}_r) = \epsilon^{0}_{i,f} +  V_{i,f}({\bf R}_{r})= \epsilon^{0}_{i,f} + \frac{2 \pi\hbar^2}{m_e} L_{i} N \int d{\bf r}_e~\rho({\bf r}_e + {\bf R}_r) ~|\Psi_{i,f}({\bf r}_e)|^2
\end{equation}
where $ \rho ({\bf R}_p)$ is the unit normalized spatial distribution function in the haze droplet at $\bf{R}_p= {\bf r}_e + {\bf R}_r $.  The mean-field potential  $V_{i,f}({\bf R}_{r})$ {induced by the haze particles} depends on the symmetry of electronic states  and  symmetry of the perturber distribution function $\rho({\bf R}_p)$. This potential can be a function of the radiator coordinate ${\bf R}_r$. The spectral intensity $I(\Delta\omega)$ is expressed via mean-filed energy shifts:
\begin{equation}
I(\Delta\omega)=  \int d^3{R}_r ~p({\bf R}_r) ~\delta ( \Delta \omega - [V_{f}({\bf R}_{r})-V_{i}({\bf R}_{r})]/\hbar )
\label{highmeanfield}
\end{equation}
where $ p({\bf R}_r)$ is the unit normalized distribution function of radiators inside and outside haze droplet. The replacement of $ p({\bf R}_r)$ with the radiator partition function can provide a temperature dependence of the radiator spatial distribution.  Angular anisotropy of the mean-field potentials $V_{i,f}({\bf R}_r)$ plays an important role in integration of the $\delta$-function in Eq.~\ref{highmeanfield}. For spherically symmetric  Born-Oppenheimer energy splitting $\Delta V(R_r) = V_{f}({\bf R}_{r})-V_{i}({\bf R}_r) $, the expression for $I(\Delta\omega)$ formally  reduces to the known approximation \citep{Allard1982}.
}
\subsubsection{Shift and broadening of spectral line in binary collisions}
The mean-filed spectral shift represents only a part of the total shift and broadening of unperturbed spectral line  with frequency $\omega_0= \Delta \epsilon_0 = \epsilon^{0}_{i} - \epsilon^{0}_f $.  Collisional effects are thermally driven, and hence temperature dependent. In the relatively dense environments, the binary collision between the radiator and a nearby perturber happens in the presence of mean-filed shifts induced by large number of other slowly moving perturbers.  The temperature-dependent binary collision broadening and shift rates  are, respectively  $\rho({\bf R}_r) \gamma(T)$, and $\rho({\bf R}_r) \sigma(T)$ \citep{Allard1982,Kotochigova2003}. 

The overall line shape is obtained upon integrating the localized contributions over the whole volume, assuming that the radiator is found at position ${\bf R}_r$ with the spherically symmetric  probability density $p({\bf R}_r)$:
\begin{equation}
I(\omega) = {1\over \pi} \int_0^\infty \frac{\rho({\bf R}_r)\gamma \; }
{[\omega - (\Delta \epsilon_0 + V_f({\bf R}_r) - V_i({\bf R}_r)) - \rho({\bf R}_r) \sigma]^2 + [\rho({\bf R}_r) \gamma]^2} \; p({{\bf R}_r)} \; d^3 R_r
\label{eq:lineform}
\end{equation}
In the above, the first term in the denominator describes the shift of the resonance frequency $\omega_0$, and the second term the broadening.

We expect  a strong transformation of the emission spectra when the radiator  is localized near  the surface {or inside haze aggregate. In such cases, the strong mean-field interaction creates  a broad emission/absorption spectra, significantly shifted from the unperturbed line at $\omega_0$.} The spectral regions with the photon frequencies  near the unperturbed spectral lines are mostly formed by collisional interactions with  shifted and broadened Lorentzian profiles \citep{Kotochigova2003,Szudy1996}, and become temperature dependent. 

The rich experimental and theoretical database  on collisional transformation of emission and absorption spectra are available for  different atomic and molecular radiators in a gas environment \citep{Szasz1985,Kotochigova2003}, where physics of shift and broadening of radiation spectra near the radiator spectral core is described by binary collisions between radiators and gas particles. Analysis of  spectral line  shift and broadening   provides valuable information on parameters of atmospheric gases and haze particles. Parameters of the collisional and mean-field  mechanisms of the shift and broadening  for different spectral lines can yield  unique vistas on physical characteristics of haze particle size, shape and material.

\section{Illustrative Results}

We demonstrate the utility of the current theoretical framework for occultation of expolanetary atmospheres, using sodium atom as the resonant radiator and argon as the haze constituent. The electron scattering length for argon has been measured from low-energy drift velocities of electrons in an H-Ar mixture, to be $L_{Ar}= -1.46$ $a_0$, \citep{Petrovic1995}, where $a_0$ is the Bohr radius. Admittedly, the haze constituency will more likely be in the form of carbon-bearing, water, and other complex molecules, but for purposes of computational efficiency, interacting sodium atoms with rare-gas atomic samples offers a convenient way to calculate many-body mean-field aspect of this theory. Below, first we describe how we calculate the binary interaction in Na-Ar. 
\subsection{Binary Potential Energy Curves}

We employ effective core potentials (ECPs), developed by \cite{nicklass:95}, to represent the 10 core electrons of both argon and sodium. The sodium core dipole polarizability is $\alpha=0.9947$ and exponential cutoff parameter $\delta = 0.62$, in atomic units,
using the core polarization potentials (CPPs) developed by \cite{fuenteabla:82} and implemented in {\sc Molpro} \citep{molpro:12}. The one-electron basis set for the valence electrons of the argon perturber is taken from \citep{nicklass:95}.
The sodium radiator is described using an extended one-electron basis set which is defined as follows:
The $s$, $p$, and $d$ orbitals are represented by a set of uncontracted 5 Gaussians with even-tempered exponents between 1.0 and 0.01 atomic units. Two sets of uncontracted Gaussian $f$ orbitals with exponents $0.08.$ and $0.008$ are also included.

We perform complete active space self-consistent field (CASSCF) calculations to obtain molecular orbitals,
which are state-averaged over the 3S ground state and 3P excited states. The active space contains the 9 valence electrons in the argon and sodium 3$s$ and 3$p$ orbitals.
Subsequently, we perform multi-reference configuration interaction (MRCI) calculations,
which includes single and double excitations from this active space.
We include the Pople size-consistency correction,
and calculate interaction energies using the counterpoise procedure of \citep{boys:70}, to correct for the basis set superposition error.
We use a dense grid in the radial coordinate, extending from $R=3$ to $20~a_0$ in steps of 0.25~$a_0$,
with additional points at 22, 25~$a_0$.
A final point at $R=50~a_0$ is used to subtract any remaining error in size consistency.
All calculations are performed using the {\sc Molpro} suite of \emph{ab initio} programs \citep{molpro:12}.

\begin{figure}
\centerline{\includegraphics[width=4.0in]{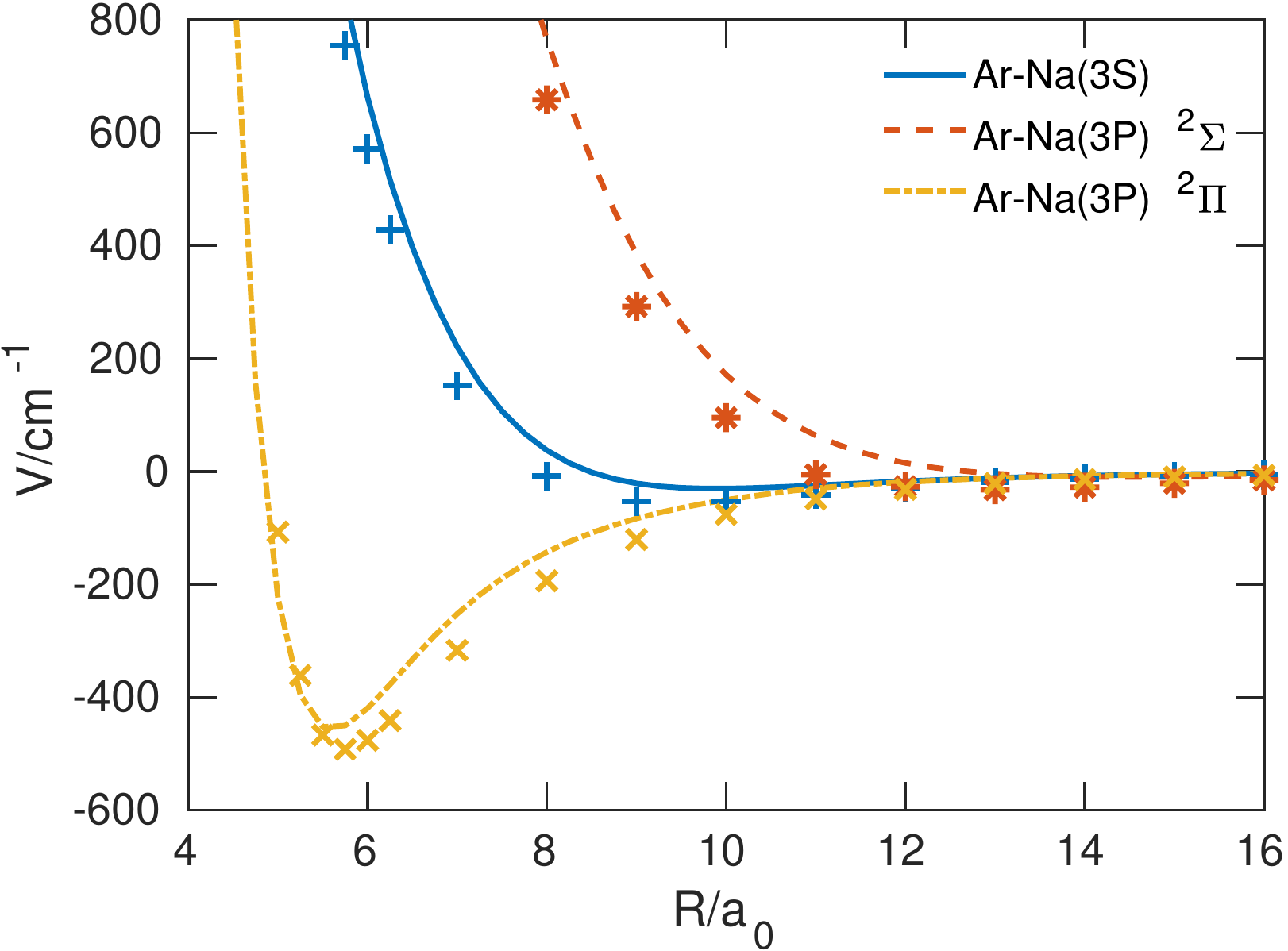}}
\caption{Potential energy curves for Ar$-$Na(3S,3P). The Na D line energy is subtracted.
Lines correspond to the potential energy curves calculated in this work, and symbols correspond to calculated potential energy values \citep{saxon:77}.
}
\label{fig:PEC}
\end{figure}

The potential energy curves are shown in Fig.~\ref{fig:PEC}.
Results obtained in this work are shown as lines.
There are two potentials correlating to the Na(3P) excited state.
The $^2\Sigma$ potential governs the interaction if the excited Na valence electron occupies a 3P orbital oriented parallel to the interatomic axis,
whereas the $^2\Pi$ potential describes excited states with the 3P orbital perpendicular to the interatomic axis.
The $^2\Pi$ excited state potential is significantly more attractive, with a well depth of around 450 cm$^{-1}$, and becomes repulsive only for much shorter interatomic separations.
The $^2\Sigma$ excited state, however, has a more repulsive potential than the ground state.
The calculated potentials are in good qualitative agreement with the results of \citet{saxon:77}.

\subsection{Shift and broadening of Na line in Ar haze}
\label{shift}
The argon droplet is modeled by a Gaussian distribution, given by a { spherically-symmetric } position dependent density $
{\rho}(R) = [N_{\rm{Ar}}/{(2\pi a^2)^{3/2}}] \exp ( - R^2/2 a^2)
$, which depends on the droplet size parameter $a$, and the number of Ar atoms, $N_{\rm{Ar}}$. A sodium atom at a distance ${R_r}$ from the center of the droplet has a shifted Lorentz absorption profile as in Eq.~\ref{eq:lineform}, where the shape of the line is determined by two contributions: a) mean field {quasi-}static contribution of neighboring argon atoms that produce the energy shifts {$\Delta V(R_r)= V_{3P}({\bf R}_r) - V_{3S}({\bf R}_r)$}, and b) collisional broadening $\Gamma$ and shift $\Sigma$ rates that are proportional to the local density: $\Gamma = \rho( R_r) \gamma$ and $\Sigma = \rho(R_{r}) \sigma$. In this model, we use the measured values $\gamma = 1.47 \times 10^{-20}$ cm$^{-1}$/cm$^{-3}$, and $\sigma = 0.75 \times 10^{-20}$ cm$^{-1}$/cm$^{-3}$ at T = 475~K \citep{Allard1982}. For T = 2000~K, we multiply the shift and broadening by a factor 1.4; this factor is obtained from a comparison of line broadening in Na-Ar collision from \citet{Jongerius1981}.

Fig.~\ref{lineprofile} illustrates the  absorption spectral line shape of the Na D line embedded in a haze environment of specific Ar density. The two cases considered, with 1000 and 10000 Ar atoms, include  haze seed particles of size given by $a = 40$ a.u. The probability of finding the sodium atom at some distance $R_{r}$ is given by the Boltzmann distribution, $p(R_r) = \frac{1}{Z}e^{-\epsilon_{3S}(R_{r})/k_BT}$, with normalization $Z = \int_0^\infty e^{-\epsilon_{3S}(R_{r})/k_BT} (4\pi R_{r}^2) dR_{r}$, $ \epsilon_{3S}(R_{r})$ {is}  the mean-field shift in the $3S$ state, see also Fig.~\ref{difference}, and $k_B$ the Boltzmann constant. In the absence of core broadening due to binary collisions, the spectral line profile will contain a characteristic discontinuity at the Na line. The line profiles are asymmetric about the radiator line center; this is an expectation of the change in sign of $\Delta \epsilon (R)$, see Fig.~\ref{difference}. {\it In situ} simulated laboratory measurements of Titan atmospheric aerosol density confirm that effective  {number densities}  of aerosol/haze particle materials are about $10^{22}$ cm$^{-3}$ for a methane haze \citep{Horst2013}, in accord with our illustrative haze particle densities.  
\begin{figure}
\includegraphics[width=3.0in]{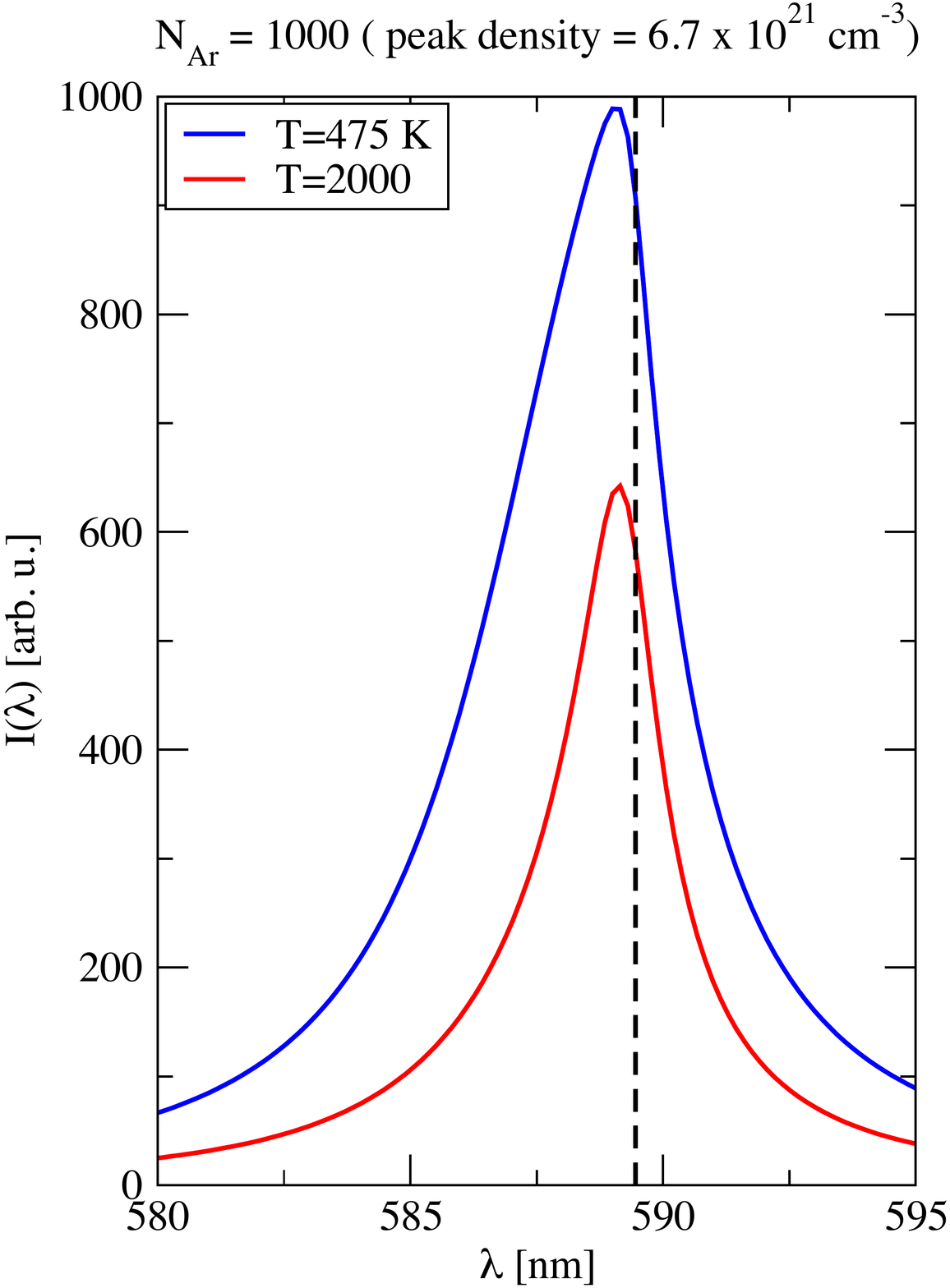}
\includegraphics[width=3.0in]{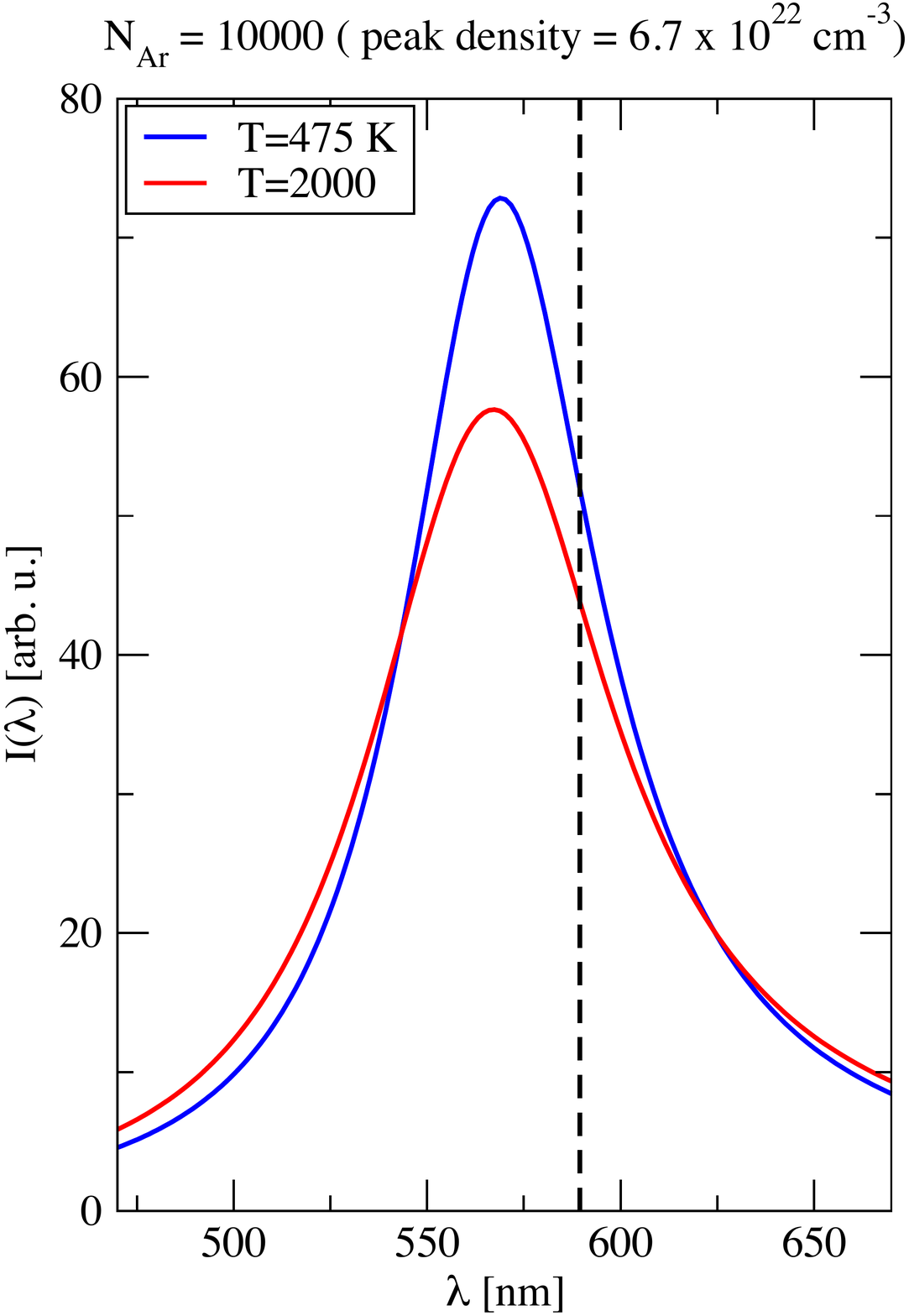}
\caption{{Temperature dependent spectral line profiles of Na D line embedded in an Ar haze of specific peak density. Near the line center, the temperature dependent is the largest due to collisional effects, and away from the center, the lines are no longer Lorentzian as the temperature-insensitive mean field interaction potentials induce asymmetric lineshapes, e.g. Fig.~\ref{difference}. }
}
\label{lineprofile}
\end{figure}
\begin{figure}
\centering{\includegraphics[width=3.2in]{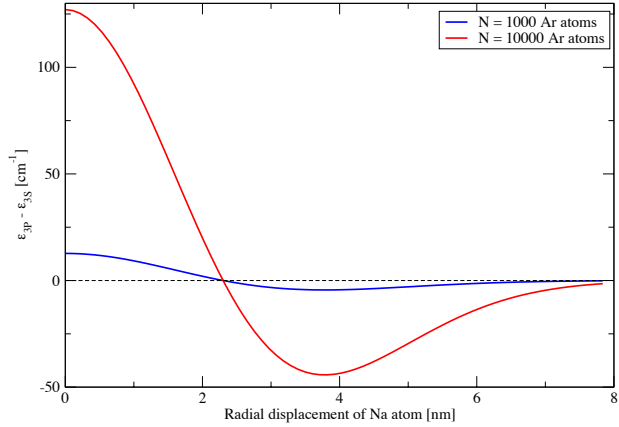}}
\caption{
Distortion of the energy difference between levels Na(3S, 3P), due to Ar haze mean field, as a function of the position of the Na atom, for two different numbers of Ar atoms ({peak} densities {$p(R_r=0)$ at $R_r=0 $}). The haze size parameter is $a=40$ a$_0$ (2.12 nm) in both cases, see Sec.~\ref{shift},  corresponding to peak densities of Ar atoms, 6.7$\times 10^{21}$~cm$^{-3}$ and 6.7$\times 10^{22}~$cm$^{-3}$ for $N = 1000$ and $N = 10000$, respectively.}
\label{difference}
\end{figure}

{\section{Summary}
In this work, a unified mean field framework for the spectral shift and broadening of an atomic radiator in a dense haze environment has been developed. Both collisional interaction, affecting the line center, and the mean field energy shifts due the presence of other atomic or molecular perturbers affecting the line wings, in hazy environments, are accounted for. The collisional contribution depends strongly on temperature and has a Lorentzian form, but the mean field shift is temperature insensitive. The line profiles in hazy environments are asymmetric and significantly broaden because the mean-field generated potential.

The current model has considerable flexibility for extension to {different} haze environments when chemical composition is mixed, particle densities vary from gaseous to solid phases, and variable porosities. The atomic radiator can be embedded within the haze, or outside and can radiate from high principal quantum numbers, where strong line shift and broadening may inhibit excitations in the first place. The collisional shift and broadening can be quantitatively calculated at various temperatures with accurate quantum mechanical methods in impact approximation \citep{Kotochigova2003}. The current model can be extended to a description of the radiator embedded into liquid or solid matrix.
 
 \section{acknowledgements}
 Z. F. was supported by an ITAMP faculty fellowship from an underrepresented institution. HRS and JFB were supported by an NSF grant to ITAMP. DV was supported by an NSF RISE grant to Texas Southern University.
\newpage

\bibliographystyle{yahapj}
\bibliography{references}

\begin{thebibliography}{}
\providecommand\natexlab[1]{#1}
\providecommand\JournalTitle[1]{#1}

\bibitem[{Allard \& Kielkopf(1982)}]{Allard1982}
Allard, N., \& Kielkopf, J. 1982,
  \href{http://dx.doi.org/10.1103/RevModPhys.54.1103}{\JournalTitle{Rev. Mod.
  Phys.}, 54, 1103}

\bibitem[{{Boatz} \& {Mario}(1994)}]{Boatz1994}
{Boatz}, J., \& {Mario}, E. 1994, \JournalTitle{J. Chem. Phys.}, 101, 3472

\bibitem[{Boys \& Bernardi(1970)}]{boys:70}
Boys, S.~F., \& Bernardi, F. 1970,
  \href{http://dx.doi.org/10.1080/00268977000101561}{\JournalTitle{Mol. Phys.},
  19, 553}

\bibitem[{{Burrows} {et~al.}(2001){Burrows}, {Hubbard}, {Lunine}, \&
  {Liebert}}]{BurrowsRMP2001}
{Burrows}, A., {Hubbard}, W.~B., {Lunine}, J.~I., \& {Liebert}, J. 2001,
  \href{http://dx.doi.org/10.1103/RevModPhys.73.719}{\JournalTitle{Reviews of
  Modern Physics}, 73, 719}

\bibitem[{{Burrows} {et~al.}(2000){Burrows}, {Marley}, \&
  {Sharp}}]{Burrows2000}
{Burrows}, A., {Marley}, M.~S., \& {Sharp}, C.~M. 2000,
  \href{http://dx.doi.org/10.1086/308462}{\JournalTitle{\apj}, 531, 438}

\bibitem[{{Burrows}(2014)}]{Burrows2014}
{Burrows}, A.~S. 2014,
  \href{http://dx.doi.org/10.1038/nature13782}{\JournalTitle{\nat}, 513, 345}

\bibitem[{{Deming} \& {Seager}(2017)}]{Seager2017}
{Deming}, D., \& {Seager}, S. 2017, \JournalTitle{ArXiv e-prints},
  \href{http://arxiv.org/abs/1701.00493}{{\sffamily arXiv:1701.00493
  [astro-ph.EP]}}

\bibitem[{{Demkov} \& {Ostrovskii}(1988)}]{Demkov1988}
{Demkov}, Y., \& {Ostrovskii}, V. 1988, {Zero-Range Potentials and Their
  Applications in Atomic Physics} (Premium Press, NY), 287

\bibitem[{ExoEU(2018)}]{ExoEU}
ExoEU. 2018, The Extrasolar Planets Encyclopaedia,
  \url{http://www.exoplanet.eu/}

\bibitem[{ExoOrg(2018)}]{ExoOrg}
ExoOrg. 2018, Exoplanet Data Explore, \url{http://www.exoplanets.org/}

\bibitem[{Fermi(1934)}]{Fermi1934}
Fermi, E. 1934, \JournalTitle{Il Nuovo Cimento (1924-1942)}, 11, 157

\bibitem[{{Fortney} {et~al.}(2003){Fortney}, {Sudarsky}, {Hubeny}, {Cooper},
  {Hubbard}, {Burrows}, \& {Lunine}}]{Fortney2003}
{Fortney}, J.~J., {Sudarsky}, D., {Hubeny}, I., {et~al.} 2003,
  \href{http://dx.doi.org/10.1086/374387}{\JournalTitle{\apj}, 589, 615}

\bibitem[{Fuentealba {et~al.}(1982)Fuentealba, Preuss, Stoll, \&
  Szentp{\'a}ly}]{fuenteabla:82}
Fuentealba, P., Preuss, H., Stoll, H., \& Szentp{\'a}ly, L.~V. 1982,
  \href{http://dx.doi.org/10.1016/0009-2614(82)80012-2}{\JournalTitle{Chem.
  Phys. Lett.}, 89, 418 }

\bibitem[{{Greenberg} \& {Li}(1999)}]{Greenberg1999}
{Greenberg}, J., \& {Li}, A. 1999, \JournalTitle{Space Sc. Rev.}, 90, 149

\bibitem[{Greene {et~al.}(2000)Greene, Dickinson, \& Sadeghpour}]{Greene2000}
Greene, C.~H., Dickinson, A.~S., \& Sadeghpour, H.~R. 2000,
  \JournalTitle{Physical Review Letters}, 85, 2458

\bibitem[{H{\"o}rst \& Tolbert(2013)}]{Horst2013}
H{\"o}rst, S.~M., \& Tolbert, M.~A. 2013, \JournalTitle{The Astrophysical
  Journal Letters}, 770, L10

\bibitem[{{H{\"o}rst} {et~al.}(2018){H{\"o}rst}, {He}, {Lewis}, {Kempton},
  {Marley}, {Morley}, {Moses}, {Valenti}, \& {Vuitton}}]{Horst2018}
{H{\"o}rst}, S.~M., {He}, C., {Lewis}, N.~K., {et~al.} 2018,
  \JournalTitle{ArXiv e-prints},
  \href{http://arxiv.org/abs/1801.06512}{{\sffamily arXiv:1801.06512
  [astro-ph.EP]}}

\bibitem[{{Jacquet} {et~al.}(2011){Jacquet}, {Zanuttini}, {Douady}, \& {et
  al.}}]{Jacquet2011}
{Jacquet}, E., {Zanuttini}, D., {Douady}, J., \& {et al.} 2011,
  \JournalTitle{J.Chem.Phys.}, 135, 174503

\bibitem[{Jongerius {et~al.}(1981)Jongerius, Bergen, Hollander, \&
  Alkemade}]{Jongerius1981}
Jongerius, M., Bergen, A.~V., Hollander, T., \& Alkemade, C. 1981,
  \href{http://dx.doi.org/https://doi.org/10.1016/0022-4073(81)90045-5}{\JournalTitle{Journal
  of Quantitative Spectroscopy and Radiative Transfer}, 25, 1 }

\bibitem[{{Kreidberg} {et~al.}(2014){Kreidberg}, {Bean}, {D{\'e}sert},
  {Benneke}, {Deming}, {Stevenson}, {Seager}, {Berta-Thompson}, {Seifahrt}, \&
  {Homeier}}]{Kreidberg2014}
{Kreidberg}, L., {Bean}, J.~L., {D{\'e}sert}, J.-M., {et~al.} 2014,
  \href{http://dx.doi.org/10.1038/nature12888}{\JournalTitle{\nat}, 505, 69}

\bibitem[{{Lenoble} {et~al.}(2013){Lenoble}, { Mishchenko}, \&
  {Herman}}]{Lenoble2013}
{Lenoble}, J., { Mishchenko}, M.~I., \& {Herman}, H. 2013, \JournalTitle{In
  Aerosol Remote Sensing (J. Lenoble et al., Eds.), pp 13-51}, 13

\bibitem[{Morley {et~al.}(2015)Morley, Fortney, Marley, Zahnle, Line, Kempton,
  Lewis, \& Cahoy}]{Morley2015}
Morley, C.~V., Fortney, J.~J., Marley, M.~S., {et~al.} 2015,
  \href{http://stacks.iop.org/0004-637X/815/i=2/a=110}{\JournalTitle{The
  Astrophysical Journal}, 815, 110}

\bibitem[{Nicklass {et~al.}(1995)Nicklass, Dolg, Stoll, \&
  Preuss}]{nicklass:95}
Nicklass, A., Dolg, M., Stoll, H., \& Preuss, H. 1995,
  \href{http://dx.doi.org/10.1063/1.468948}{\JournalTitle{J. Chem. Phys.}, 102,
  8942}

\bibitem[{{Ortiz} {et~al.}(1996){Ortiz}, {Moreno}, \& {Molina}}]{Ortiz1996}
{Ortiz}, J.~L., {Moreno}, F., \& {Molina}, A. 1996, \JournalTitle{Icarus}, 119,
  53

\bibitem[{Petrovic {et~al.}(1995)Petrovic, O'Malley, \&
  Crompton}]{Petrovic1995}
Petrovic, Z.~L., O'Malley, T.~F., \& Crompton, R.~W. 1995,
  \JournalTitle{Journal of Physics B: Atomic, Molecular and Optical Physics},
  28, 3309

\bibitem[{{Pont} {et~al.}(2008){Pont}, {Knutson}, {Gilliland}, {Moutou}, \&
  {Charbonneau}}]{Pont2008}
{Pont}, F., {Knutson}, H., {Gilliland}, R.~L., {Moutou}, C., \& {Charbonneau},
  D. 2008,
  \href{http://dx.doi.org/10.1111/j.1365-2966.2008.12852.x}{\JournalTitle{\mnras},
  385, 109}

\bibitem[{{Pont} {et~al.}(2013){Pont}, {Sing}, {Gibson}, {Aigrain}, {Henry}, \&
  {Husnoo}}]{Pont2013}
{Pont}, F., {Sing}, D.~K., {Gibson}, N.~P., {et~al.} 2013,
  \href{http://dx.doi.org/10.1093/mnras/stt651}{\JournalTitle{\mnras}, 432,
  2917}

\bibitem[{{Rannou} {et~al.}(2002){Rannou}, {Hourdin}, \& {McKay}}]{Rannou2002}
{Rannou}, P., {Hourdin}, F., \& {McKay}, C. 2002, \JournalTitle{Nature}, 418,
  853

\bibitem[{Saxon {et~al.}(1977)Saxon, Olson, \& Liu}]{saxon:77}
Saxon, R.~P., Olson, R., \& Liu, B. 1977,
  \href{http://dx.doi.org/10.1063/1.435183}{\JournalTitle{J. Chem. Phys.}, 67,
  2692}

\bibitem[{{Sciamma-O'Brien} {et~al.}(2017){Sciamma-O'Brien}, {Upton}, \&
  {Salama}}]{Sciamma2017}
{Sciamma-O'Brien}, E., {Upton}, K., \& {Salama}, F. 2017,
  \JournalTitle{Icarus}, 289, 214

\bibitem[{{Seager} \& {Deming}(2010)}]{Seager2010}
{Seager}, S., \& {Deming}, D. 2010,
  \href{http://dx.doi.org/10.1146/annurev-astro-081309-130837}{\JournalTitle{\araa},
  48, 631}

\bibitem[{{Seager} \& {Sasselov}(2000)}]{Seager2000}
{Seager}, S., \& {Sasselov}, D.~D. 2000,
  \href{http://dx.doi.org/10.1086/309088}{\JournalTitle{\apj}, 537, 916}

\bibitem[{{Seignovert} {et~al.}(2017){Seignovert}, {Rannoua}, {Lavvasa}, \& {et
  al.}}]{Seignovert2017}
{Seignovert}, B., {Rannoua}, P., {Lavvasa}, P., \& {et al.} 2017,
  \JournalTitle{Icarus}, 292, 13

\bibitem[{{Sing} {et~al.}(2011){Sing}, {D{\'e}sert}, {Fortney}, {Lecavelier Des
  Etangs}, {Ballester}, {Cepa}, {Ehrenreich}, {L{\'o}pez-Morales}, {Pont},
  {Shabram}, \& {Vidal-Madjar}}]{Sing2011}
{Sing}, D.~K., {D{\'e}sert}, J.-M., {Fortney}, J.~J., {et~al.} 2011,
  \href{http://dx.doi.org/10.1051/0004-6361/201015579}{\JournalTitle{\aap},
  527, A73}

\bibitem[{Spake {et~al.}(2018)Spake, Sing, Evans, Oklop{\v c}i{\'c}, Bourrier,
  Kreidberg, Rackham, Irwin, Ehrenreich, Wyttenbach, Wakeford, Zhou, Chubb,
  Nikolov, Goyal, Henry, Williamson, Blumenthal, Anderson, Hellier,
  Charbonneau, Udry, \& Madhusudhan}]{Spake2018}
Spake, J.~J., Sing, D.~K., Evans, T.~M., {et~al.} 2018, \JournalTitle{Nature},
  557, 68

\bibitem[{{Szasz}(1985)}]{Szasz1985}
{Szasz}, L. 1985, {Pseudopotential theory of atoms and molecules} (John Wiley
  and Sons, Inc., New York)

\bibitem[{{Szudy} \& { Baylis}(1996)}]{Szudy1996}
{Szudy}, J., \& { Baylis}, W. 1996, \JournalTitle{Physics Reports}, 266, 127

\bibitem[{{Tinetti} {et~al.}(2007){Tinetti}, {Vidal-Madjar}, {Liang},
  {Beaulieu}, {Yung}, {Carey}, {Barber}, {Tennyson}, {Ribas}, {Allard},
  {Ballester}, {Sing}, \& {Selsis}}]{Tinetti2007}
{Tinetti}, G., {Vidal-Madjar}, A., {Liang}, M.-C., {et~al.} 2007,
  \href{http://dx.doi.org/10.1038/nature06002}{\JournalTitle{\nat}, 448, 169}

\bibitem[{{Tomasko} {et~al.}(1986){Tomasko}, {Karkoschka}, \&
  {Martinek}}]{Tomasko1986}
{Tomasko}, M.~G., {Karkoschka}, J., \& {Martinek}, S. 1986,
  \JournalTitle{Icarus}, 65, 218

\bibitem[{Vrinceanu {et~al.}(2004)Vrinceanu, Kotochigova, \&
  Sadeghpour}]{Kotochigova2003}
Vrinceanu, D., Kotochigova, S., \& Sadeghpour, H.~R. 2004,
  \href{http://dx.doi.org/10.1103/PhysRevA.69.022714}{\JournalTitle{Phys. Rev.
  A}, 69, 022714}

\bibitem[{Werner \& {P. J. Knowles \em et al.}(2012)}]{molpro:12}
Werner, H.-J., \& {P. J. Knowles \em et al.} 2012, {\sc molpro}: a package of
  ab initio programs, version 2012.1

\bibitem[{{West}(1991)}]{West1991}
{West}, R. A.and~{Smith}, P.~H. 1991, \JournalTitle{Icarus}, 90, 330

\bibitem[{{Wong} {et~al.}(2003){Wong}, {Yung}, \& {Friedson}}]{Wong2003}
{Wong}, A., {Yung}, Y.~L., \& {Friedson}, A.~J. 2003,
  \JournalTitle{Geophys.Res.Lett}, 30, 8

\bibitem[{Zhang {et~al.}(2017)Zhang, Strobel, \& Imanaka}]{zhang2017}
Zhang, X., Strobel, D.~F., \& Imanaka, H. 2017,
  \href{http://dx.doi.org/10.1038/nature24465}{\JournalTitle{Nature}, 551, 352
  EP }

\end{thebibliography}

\end{document}